\documentclass[12pt]{article}

\usepackage[utf8]{inputenc}
\usepackage{amsmath}
\usepackage{amsfonts}
\usepackage{amssymb}
\usepackage{graphicx}
\usepackage{float}
\usepackage[font={small}]{caption}
\usepackage{subcaption}
\usepackage{subfig}
\usepackage{url}
\usepackage{cite}

\usepackage{color}
\usepackage{calc}
\usepackage[english]{babel}
\usepackage[T1]{fontenc}
\usepackage{graphicx}
\usepackage{verbatim}
\usepackage{enumerate}

\def\Bmp#1{ \begin{minipage}{#1} }
\def\Bmpc#1{ \begin{minipage}[c]{#1} }
\def\Bmpt#1{ \begin{minipage}[t]{#1} }
\def\Bmpb#1{ \begin{minipage}[b]{#1} }
\def\Emp{ \end{minipage} }

\def\C{{\mathcal{C}}}

\def\O{{\mathcal{O}}}

\def\Z{{\mathcal{Z}}}

\def\tf0{\tilde{\varphi}_{0}}

\def\CC{{\mathbb{C}}}

\def\RR{{\mathbb{R}}}

\def\n{{\bf n}}

\def\x{{\bf x}}

\def\u{{\bf u}}
\def\0{{\bf 0}}

\begin{document}

\title{Drift Due to Two Obstacles in Different Arrangements}
\author{Sergei Melkoumian$^1$ and Bartosz Protas$^2$
\\ \\
$^1$School of Computational Science and Engineering, \\
McMaster University  \\
Hamilton, Ontario L8S4K1, CANADA 
\\ \\
$^2$Department of Mathematics \& Statistics, \\
McMaster University  \\
Hamilton, Ontario L8S4K1, CANADA }

\date{\today}
\maketitle

\begin{abstract}
{ We study the drift induced by the passage of two
    cylinders through an unbounded extent of inviscid incompressible
    fluid under the assumption that the flow is two-dimensional and
    steady in the moving frame of reference. The goal is to assess how
    the resulting total particle drift depends on the parameters of
    the geometric configuration, namely, the distance between the
    cylinders and their angle with respect to the direction of
    translation. This problem is studied by numerically computing, for
    different cylinder configurations, the trajectories of particles
    starting at various initial locations. The velocity field used in
    these computations is expressed in closed form using methods of
    the complex function theory and the accuracy of calculations is
    carefully verified.  We identify cylinder configurations which
    result in increased and decreased drift with respect to the
    reference case when the two cylinders are separated by an infinite
    distance. Particle trajectories shed additional light on the
    hydrodynamic interactions between the cylinders in configurations
    resulting in different drift values.  This ensemble of results
    provides insights about the accuracy of models used to study
    biogenic {transport}.}
\end{abstract}

\begin{flushleft}
\end{flushleft}


\section{Introduction}
\label{sec:intro}

Drift is a phenomenon whereby fluid particles experience a net
displacement after the passage of an obstacle in an unbounded domain.
It has recently gained renewed attention in {connection with
  transport and mixing} in the oceans caused by swimming organisms
where drift may play an important role in the overall energy transfer
in the oceans \cite{katija2009}. Other applications involve multiphase
flows \cite{e03a} and we refer the reader to our recent study
\cite{melkoumian2014} for a survey of the relevant literature.  As was
initially introduced in the classic studies by Maxwell and Darwin
\cite{maxwell1870, darwin1953}, most investigations of drift rely on
potential models of flows around a circular cylinder. The effects of
wake vortices on drift were investigated in \cite{melkoumian2014}
based on the potential flow models of F\"oppl and Kirchhoff. It was
found that, for the F\"oppl wake model \cite{f13}, the effect of the
wake vortices was to increase the total drift, except for very small
wake sizes for which the drift was actually decreased.  On the other
hand, for the Kirchhoff model \cite{lc07} the total drift is
unbounded.

\begin{figure}
\centering
\includegraphics[width=0.5\textwidth]{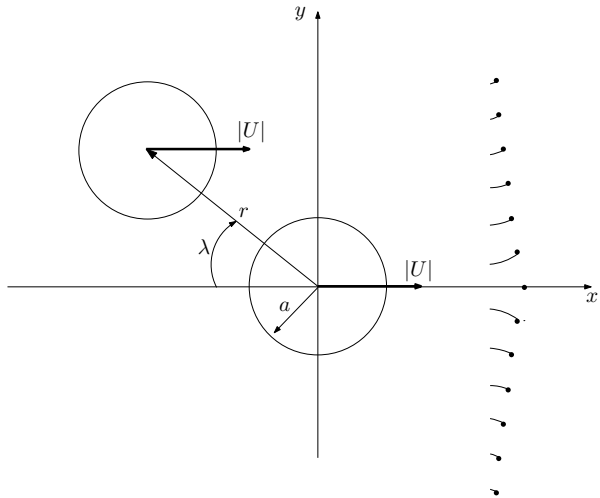}
\caption{Schematic of the flow configuration {at time $t=0$ in
    the laboratory frame of reference}.}
\label{fig:setup}
\end{figure}

{Information about particle displacements resulting from the
  passage of an obstacle, or obstacles, can be used to characterize
  net transport, when these displacements are integrated to produce
  total drift \cite{childress2009}, or mixing, when the squared
  displacements are averaged to describe the effective diffusivity
  \cite{t15}. In the present investigation we are concerned with the
  former quantity.}  A typical approach \cite{tc10a} is to view the
effects of multiple objects as independent of each other, which is
equivalent to assuming that the drift induced by the passage of
multiple objects is a simple sum of the drifts induced by the
individual bodies. In other words, in this approach the geometric
nonlinearity of the hydrodynamic interactions is neglected by assuming
the drift to be linearly additive. The goal of this study is to
address this issue in more detail by considering how the actual drift
induced by the passage of multiple objects depends on their geometric
configuration. {In the context of mixing, information about the
  displacements of individual particles was used to determine the
  effective diffusivity in \cite{tc10a, ltc11, lz14}. In particular,
  the study \cite{lz14} examined the effect of the passage of multiple
  objects in different geometric configurations.

  In our study} we {focus} on an idealized flow problem in which
two identical circular cylinders pass through an unbounded extent of
fluid such that their geometric configuration remains unchanged. As a
results of this simplification, there are only two parameters in the
problem, namely, the distance $r$ between the centers of the obstacles
and their inclination, measured by the angle $\lambda$, with respect
to the direction of motion (Figure \ref{fig:setup}). {Due to symmetry,
  we will only need to consider the values $0 \le \lambda \le \pi/2$.}
It will be demonstrated that the resulting total drift may be
increased or decreased with respect to the linearly additive case
depending on the values of the parameters $r$ and $\lambda$. The
values of the drift obtained in various geometric configurations are
correlated with the patterns exhibited by the particle trajectories in
the different cases, thereby offering physical insights into the
kinematic mechanisms underlying the increase or decrease of drift.

The structure of the paper is as follows: in the next section we
define precisely drift and discuss various ways of computing it in
general flows; in Section \ref{sec:models} we restrict our attention
to two-dimensional (2D) potential flows induced by translating
objects; next, in Section \ref{sec:numer} we introduce and validate
our numerical approach and then in Section \ref{sec:results} we
present and discuss the computational results; finally, conclusions are
deferred to Section \ref{sec:final}.

\section{Drift: Definition and Calculation}
\label{sec:drift}

We will consider the motion of $N$ circular cylinders with boundaries
$\Gamma_i$, $i = 0, 1,\dots, N-1$, passing through an incompressible
inviscid fluid of unit density in a 2D unbounded domain $\Omega$.
{In our analysis we will use two coordinate systems: one
  associated with the laboratory frame of reference and the other
  attached to the moving obstacles (in the latter case we will assume
  that the cylinder with the boundary $\Gamma_0$ has its center at the
  origin).} For a point $(x,y) \in \Omega$, the position and velocity
vectors {in the laboratory frame of reference} will be denoted,
respectively, $\x = [x, y]^T$ and $\u(\x) = [u_x, u_y ]^T$, where
$u_x$ and $u_y$ are the $x$ and $y$ components.  It is assumed that
the obstacles pass {from} $x = -\infty$ to $x = \infty$ with a constant
speed and in such a way that the distance $r$ between the obstacle
centers and the inclination angle $\lambda$ remain unchanged (Figure
\ref{fig:setup}). In the moving frame of reference, we {will}
denote the position and velocity by $\x' = [x', y']^T$ and $\u'(\x') =
[u_x', u_y' ]^T$. {It is assumed that in this frame of
  reference} the flow is steady and potential, satisfies the no
through-flow boundary condition $\u'\cdot\n=0$ on each cylinder
boundary $\Gamma_i$, $i=0,\dots,N-1$, and approaches the uniform
stream at infinity $\u'(\x') \rightarrow U \hat{\x}'$ as $|\x'|
\rightarrow \infty$, where $U = -1$ and $\hat{\x}'$ is the unit vector
associated with the $x'$-axis. {We remark that, since the
  obstacles translate in the direction of the $x$-axis, the $y$
  coordinates in the two coordinate systems coincide ($y \equiv y'$).}
While the set-up described certainly represents a highly idealized
configuration (especially the aspect concerning an infinite passage
time), due to the geometric nonlinearity of the problem it allows us
to characterize the interactions between obstacles and advected
particles at a fundamental level.

Drift is defined in terms of the trajectories of individual particles
displaced as the obstacles move through the fluid.  Let the initial
position of the particle at $t=0$ be $\x_0$ and $[x(t;\x_0),
y(t;\x_0)]^T$ denote the corresponding particle trajectory for $t<0$
and $t>0$. Then, the drift of the particle initially at $\x_0$ is
defined as 
\begin{equation} 
\xi(\x_0) := \int_{-\infty}^{\infty} u_x(x(t;\x_0), y(t;\x_0))\,dt
\label{eq:xi}
\end{equation}
(the symbol ``$:=$'' defines the quantity on the left-hand side with
the quantity on the right-hand side).  To measure the amount of
{fluid transport} induced by the obstacles, the quantity that we are most interested in
is the {\em total drift area} $D$ representing the integral
displacement of all particles initially located on a line
perpendicular to the path of the obstacles at an infinite distance
upstream {in the moving frame of reference}, i.e.,
\begin{equation}
D := \int_{-\infty}^{\infty} \xi(y_{\infty})\, dy_{\infty} =  \int_{-\infty}^{+\infty}{\xi(\psi') \, d\psi'},
\label{eq:D}
\end{equation}
where $y_{\infty}$ is the transverse coordinate of the particle's
position when $t \rightarrow -\infty$ {(which is the same in
  both coordinate systems) and $\psi'$ is the streamfunction in the
  moving frame of reference} (with a slight abuse of notation, $\xi$
may be equivalently considered a function of $\x_0$, $y_{\infty}$ or
$\psi$). The two integrals in \eqref{eq:D} are equal, because
${\psi'} \rightarrow y_{\infty}$ as ${x'} \rightarrow
\infty$ which is a consequence of the far-field boundary condition
satisfied by the velocity field. The total drift area $D$ involves two
nested improper integrals (in expressions \eqref{eq:xi} and
\eqref{eq:D}) and this quantity is well-defined only if the order of
integration is as indicated here, i.e., first with respect to time
(or, equivalently, the streamwise coordinate) and then with respect to
the transverse coordinate \cite{yih1985,benjamin1986,yih1997,ebh94a}.

In general, as outlined in \cite{melkoumian2014}, there are two
effective ways to evaluate the total drift area $D$ numerically. In
the first method, one can use a suitably transformed definition
formula \eqref{eq:D} combined with the particle displacement given in
\eqref{eq:xi}. The second method is to evaluate the total drift area
by using Darwin's theorem \cite{darwin1953} which stipulates that $D
= M$, where $M$ is the added mass and the fluid density is assumed
equal to the unity. In the case of a single obstacle ($N=1$), the
added mass can be evaluated as follows \cite{s92}
\begin{equation}
M = \oint \limits_\C  \phi n_x \, ds,
\label{eq:M}
\end{equation}
where the contour $\C$ is the boundary of the largest simply-connected
region with closed streamlines. Generalization of this approach to the
case with multiple boundaries ($N\ge 1$) is straightforward. In the
present study we will follow the first approach which was thoroughly
validated in \cite{melkoumian2014}, as it has the additional advantage
of providing the particle trajectories, thereby offering insights
about various kinematic mechanisms at play.

From the practical point of view, the most convenient way to evaluate
the improper integral \eqref{eq:xi} is to set the initial particle
positions $\x_0$ at $t=0$ and then obtain the trajectories by
integrating the system
\begin{equation}
\frac{d \x(t)}{dt} = \u(\x(t)), \quad \x(0) = \x_0
\label{eq:dxdt}
\end{equation}
forward and backward in time, i.e., for $t \rightarrow \pm \infty$,
for different $\x_0$. Since in the moving frame of reference the
initial particle positions in formula \eqref{eq:D} are given for
${x'} \rightarrow \infty$, they need to be transformed to
positions with finite streamwise locations {in the laboratory
  frame}, e.g., $\x_0 = [0, y_0]^T$.  Since for a particle on a given
streamline, the streamfunction ${\psi'}$ is constant and equal
to some $C$, we have
\begin{equation}
C = {\psi'(0,y_0) = \lim_{x' \rightarrow \infty} \psi(x',y_{\infty})} = y_{\infty}. 
\label{eq:C}
\end{equation}
Defining $g(y_0) := \psi(0,y_0) = y_{\infty}$ as the map between the
$y$-coordinates of the particle at ${x'} = 0$ and at ${x'}
= \infty$, we obtain
\begin{equation}
\frac{d y_{\infty}}{d y_0} = {\dot{g}}(y_0),
\label{eq:dg}
\end{equation}
{where the dot denotes differentiation,} so that \eqref{eq:D}
becomes
\begin{align}
D 	&= \int_{-\infty}^{0} {\xi(\psi') \, d\psi'} 
         + \int_{0}^{+\infty} {\xi(\psi') \, d\psi'}, \nonumber \\ 
	&= \int_{-\infty}^{-1}\xi(g(y_0))\, {\dot{g}}(y_0) \, dy_0 
         + \int_{+1}^{\infty}\xi(g(y_0))\, {\dot{g}}(y_0) \, dy_0.
\label{eq:D2}
\end{align}
The upper bound in the first integral on the right-hand side (RHS) in
\eqref{eq:D2} and the lower bound in the second integral are now equal
to $-1$ and $+1$, respectively, because the particle on the streamline
with ${\psi'} = 0$ has the {initial} coordinate $y_0 =
\pm1$ at $x_0 = 0$.  {The reason is that the value of the
  streamfunction} on the streamline which in the moving frame of
reference coincides with the boundary $\Gamma_0$ of the first obstacle
{may be chosen as $\psi' = 0$.}

\section{Flow Model}
\label{sec:models}

First, in Section \ref{sec:mult-cylinderflow}, we will briefly review
the potential flow theory for the general case of $N$ cylinders with
arbitrary radii and positions. Then, in Section
\ref{sec:two-cylinderflow}, we will restrict this description to the
case of two identical circular cylinders of unit radius ($a = 1$).
This is the case for which we will provide computational results and
discussion in the remainder of the paper.

Hereafter, without the risk of confusion, we will interchangeably use
the vector and complex notation for various vector quantities. A point
$\x' = [x', y']^T$ in the moving frame of reference will therefore
also be represented as $z = x'+iy'$, where $i := \sqrt{-1}$, and the
fluid velocity $\u'(\x') = [u_x', u_y']^T$ as $V(z) =
(u_x'-iu_y')(z)$.  Since the velocity field is assumed incompressible
and irrotational, it will be expressed in terms of the complex
potential $W(z) = ({\phi' + i\psi'})(z)$ as $V(z) = dW / dz$,
where ${\phi'}$ and ${\psi'}$ are, respectively, the
scalar potential and the streamfunction. {Since the flow models
  are defined in the moving frame of reference, in order to simplify
  the notation, we will drop the primes from the quantities defined in
  the complex plane.}

\subsection{Potential Flows Past Multiple Cylinders}
\label{sec:mult-cylinderflow}

To determine the complex potential and velocity in a flow past
multiple cylinders, we apply the methods based on the Schottky-Klein
prime function described in \cite{c06a}. They rely on the definition of
suitable conformal maps.  The set-up of the problem is such that, in
addition to the cylinder with the boundary $\Gamma_0$ located at the
origin, there are $N-1$ cylinders with centers and radii denoted
{$\left\{D_j, Q_j \right\}_{j=1}^{N-1}$} in the $z$-plane and
{$\left\{\delta_j, q_j \right\}_{j=1}^{N-1}$} in the $\zeta$-plane,
where {$Q_j,q_j \in \RR$} and {$D_j,\delta_j \in \CC$}, $j=1,\dots,N-1$
(it is assumed that the cylinders have no points of contact).  We
define a conformal map
\begin{equation}
\zeta(z) = \frac{a}{z - b},
\label{eq:zeta}
\end{equation}
where $a \in \RR$ and $b \in \CC$ are the radius and position of the
cylinder $\Gamma_0$ in the $z$-plane (since for this cylinder we have
$a=1$ and $b=0$, map \eqref{eq:zeta} simplifies to $\zeta(z) = 1 / z$).
With this map, we have that the point $\beta = 0$ in the $\zeta$-plane
maps to the point $z(\beta) = \infty$ in the $z$-plane. The relations
between the positions and radii of the remaining cylinders in the $z$
and $\zeta$ planes are {then} given by
\begin{subequations}
\label{eq:DQ}
\begin{align}
D_j = \frac{\bar{\delta_j}}{|\delta_j|^2 - q_j^2}, \\
Q_j = \frac{q_j}{|\delta_j|^2 - q_j^2}.
\end{align}
\end{subequations}
Next, for $j = 1,\dots, N-1$, we define the M\"obius maps 
\begin{align}
\theta_j(\zeta) = \frac{a_j\zeta + b_j}{c_j\zeta + d_j},
\end{align}
where
\begin{equation*}
a_j = q_j - \frac{|\delta_j|^2}{q_j}, \qquad b_j = \frac{\delta_j}{q_j}, \qquad
c_j = -\frac{\bar{\delta_j}}{q_j}, \qquad d_j = \frac{1}{q_j}.
\end{equation*}
Then, the Schottky-Klein prime function is given by 
\begin{equation}
\omega(\zeta, \gamma) = (\zeta - \gamma) \tilde{\omega}(\zeta, \gamma),
\label{eq:SK}
\end{equation}
where
\begin{equation*}
\tilde{\omega}(\zeta, \gamma) = 
\prod_{\theta_k} \frac{ (\theta_k(\zeta) - \gamma)(\theta_k(\gamma) - \zeta)}{(\theta_k(\zeta) - \zeta)(\theta_k(\gamma) - \gamma)}
\end{equation*}
in which the product is taken over the composition of all possible
M\"obius maps, called the Schottky group \cite{crowdy2007}. Function
\eqref{eq:SK} can be evaluated using an algorithm involving
a Fourier-Laurent expansion described in \cite{crowdy2007}. The complex
potential characterizing the potential flow past a system of $N$
cylinders subject to the no through-flow boundary conditions imposed on
their boundaries is then given by
\begin{equation}
W(\zeta, \beta) = Ua \left( \frac{\partial}{\partial \bar{\gamma}} - \frac{\partial}{\partial \gamma} \right) 
\log{ \left[ \frac{\omega(\zeta, \gamma)}{|\gamma| \omega( \zeta, \bar{\gamma}^{-1} ) } \right] } \bigg|_{\gamma = \beta}
\label{eq:W_Sk}
\end{equation}
and the corresponding velocity can be computed as $V(z) = (dW/d\zeta)
(d\zeta/dz)$. The terms with derivatives with respect to $\gamma$,
$\bar{\gamma}^{-1}$ and $\zeta$ may be computed numerically using
finite differences. Alternatively, noting that $\tilde{\omega}(\zeta, \gamma)
= \tilde{\omega}(\gamma, \zeta)$, the terms with derivatives with respect to
$\gamma$ and $\bar{\gamma}^{-1}$ may also be computed by
differentiating the Fourier-Laurent expansion of $\tilde{\omega}(\gamma,
\zeta)^2$, cf.~\cite{crowdy2007}.

\subsection{Potential Flow Past Two Cylinders}
\label{sec:two-cylinderflow}

\begin{figure}
\centering
\includegraphics[width=1\textwidth]{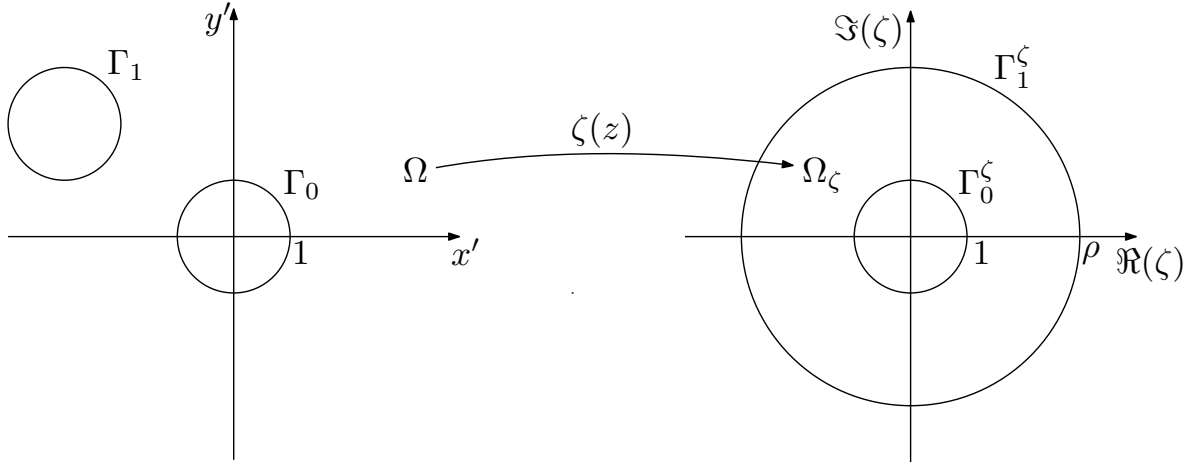}
\caption{The conformal map \eqref{eq:zeta2} from the physical plane
  $z$ {(corresponding to the moving frame of reference)} to the
  $\zeta$ plane.}
\label{fig:Map}
\end{figure}

Although it is possible to adopt the techniques from Section
\ref{sec:mult-cylinderflow} to find the complex potential for the flow
past two cylinders, a formulation is available which involves elliptic
functions \cite{johnson2004}. In fact, the two approaches are
equivalent, since in the case of two cylinders, the complex potential
\eqref{eq:W_Sk} involving the Schottky-Klein prime function reduces to
that involving elliptic functions \cite{c06a}. {For two
  cylinders, the simplified Laurent series for the Schottky-Klein
  prime function is given in\cite{Crowdy2010}.} {While this is a
  viable approach, in the present study we choose to employ a
  formulation based on elliptic functions which is described below.
  The reason is that in the course of extensive tests we performed
  this formulation was in fact found to possess better computational
  properties in terms of susceptibility to truncation and round-off
  errors than the more general approach described in Section
  \ref{sec:mult-cylinderflow}, especially, for extreme values of
  problem parameters.}

We start by defining the conformal map
\begin{equation}
\zeta(z) = \frac{\alpha e^{i \lambda} z - 1}{e^{i \lambda} z - \alpha},
\label{eq:zeta2}
\end{equation}
where $\alpha = e^{\beta}$ and $\beta = \cosh^{-1}{(-r/2)}$.  As in
\cite{johnson2004}, the unit cylinder (with boundary $\Gamma_0$)
{located} at the origin in the $z$-plane maps to a unit cylinder
at the origin in the $\zeta$-plane (with boundary $\Gamma^{\zeta}_0$),
cf.~Figure \ref{fig:Map}. The second cylinder $\Gamma_1$ is then
mapped to cylinder $\Gamma^{\zeta}_1$ with the center at the origin
and with radius $\rho > 1$, where
\begin{equation*}
\rho = e^{\gamma} \quad \text{and} \quad \gamma = \cosh^{-1}[(r^2-2)/2].
\end{equation*}
Given the conformal map \eqref{eq:zeta2} and defining $\tau :=
\log{\zeta}$, the complex potential of the flow past two cylinders
becomes
\begin{equation}
W(\tau) = iU \left[ e^{-i \lambda} \Z(i (\tau - \beta)) - e^{i \lambda} \Z(i (
\tau + \beta)) - \frac{ 2 \Z (\pi) \sin{\lambda} }{\pi} \tau \right],
\label{eq:W}
\end{equation}
where $\Z : \mathbb{C} \rightarrow \mathbb{C}$ is the Weierstrass zeta
function \cite{Olver:2010:NHMF}. To obtain the velocity field $V(z)$,
we use the relation $\wp (z) = -\dot{\Z}(z)$, where $\wp : \mathbb{C}
\rightarrow \mathbb{C}$ is the Weierstrass P function
\cite{Olver:2010:NHMF} {and the dot denotes differentiation with
  respect to $z$.} Applying the chain rule, we obtain
\begin{align}
V(z) &= \frac{dW}{dz} = \frac{dW}{d \tau}\frac{d \tau}{d \zeta}\frac{d
\zeta}{dz}, \nonumber \\
&= U \left[ e^{-i \lambda} \wp(i (\tau - \beta)) - e^{i \lambda} \wp(i (
\tau + \beta)) - \frac{ 2 \Z (\pi) \sin{\lambda} }{\pi} \right] \frac{1}{\zeta} \frac{e^{i \lambda} (1 - \alpha^2)}{(e^{i \lambda} z
- \alpha)^2} .
\label{eq:V}
\end{align}
Standard definitions of the special functions $\Z$ and $\wp$ are given
below \cite{Olver:2010:NHMF}: if $\nu_1$ and $\nu_2$ are nonzero
complex numbers such that $\Im (\nu_2 / \nu_1 ) > 0$, then the
set of points $w = 2m \nu_1 + 2n \nu_2$ with $m, n \in
\mathbb{Z}$ constitutes a lattice $\mathbb{L}$ with generators
$2\nu_1$ and $2\nu_2$. Then, the Weierstrass zeta and
P functions are defined as
\begin{subequations}
\begin{align}
\Z(z) &= \frac{1}{z^2} + 
\sum_{w \in \mathbb{L} \backslash \{ 0 \}} \left[ \frac{1}{(z-w)^2} - \frac{1}{w^2} \right], \label{eq:W-Z}\\
\wp(z) &= \frac{1}{z} + 
\sum_{w \in \mathbb{L} \backslash \{ 0 \}} \left( \frac{1}{z-w} + \frac{1}{w} + \frac{z}{w^2} \right), \label{eq:W-p}
\end{align}
\end{subequations}
respectively. Following \cite{johnson2004}, we set the half-periods to
be $\nu_1 = \pi$ and $\nu_2 = i \gamma$, so that we have for $j =
1, 2$,
\begin{align*} 
\Z(z + 2\nu_j) &= \Z(z) + 2\Z(\nu_j), \\
\wp(z + 2\nu_j) &= \wp(z)
\end{align*}
and thus $\Z$ and $\wp$ are, respectively, a quasi-periodic and
periodic functions on the lattice $\mathbb{L}$.  Numerical evaluation
of the Weierstrass zeta and P functions using the standard definitions
in \eqref{eq:W-Z} and \eqref{eq:W-p} is impractical, because of the
slow convergence of the summations with respect to the lattice points
$w \in \mathbb{L} \backslash \{ 0 \}$.  An alternate formulation
expresses the Weierstrass zeta and P functions in terms of the Jacobi
theta functions which have Fourier series that do converge rapidly
\cite{Olver:2010:NHMF}. For some given $q = e^{i \pi \nu_2 /
  \nu_1}$, we then have
\begin{subequations}
\begin{align}
\Z(z) &= \frac{\pi}{2 \nu_1} \frac{d \ln \theta_1 (\pi z / (2 \nu_1), q)}{d(\pi z / (2 \nu_1))} -\frac{\pi^2}{12 \nu_1 ^2} \frac{\dddot{\theta}_1(0, q)}{\dot{\theta}_1 (0, q)} z, \label{eq:W-Z2}\\
\wp(z) &=\left[ \frac{\pi \theta_3(0, q) \theta_4(0, q) \theta_2(\pi z / (2\nu_1), q)}{2 \nu_1 \theta_1(\pi z / (2 \nu_1), q)} \right]^2 + \frac{\pi^2}{12 \nu_1 ^2} \left( \theta_2 ^4 (0, q) + 2 \theta_4 ^ 4 (0, q) \right), \label{eq:W-p2}
\end{align}
\end{subequations}
where the Jacobi theta functions $\theta_j$, $j = 1, 2, 3, 4$, are
given by the Fourier series
\begin{subequations}
\begin{align}
\theta_1(z, q) &= 2 \sum_{n=0}^{\infty} (-1)^n q^{(n + \frac{1}{2})^2} \sin((2n+1)z), \label{eq:Th1}\\
\theta_2(z, q) &= 2 \sum_{n=0}^{\infty} q^{(n + \frac{1}{2})^2} \cos((2n+1)z), \label{eq:Th2}\\
\theta_3(z, q) &= 1 + 2 \sum_{n=1}^{\infty} q^{n^2} \cos(2nz), \label{eq:Th3}\\
\theta_4(z, q) &= 1 + 2 \sum_{n=1}^{\infty} (-1)^n q^{n^2} \cos(2nz) \label{eq:Th4}
\end{align}
\end{subequations}
{and dots represent (repeated) differentiation with respect to
  $z$.}  The derivative terms $\dot{\theta}_1$ and $\dddot{\theta}_1$
appearing in \eqref{eq:W-Z2} may be computed by differentiating
\eqref{eq:Th1}.  For points $z$ in the parts of the domain $\Omega$
that we are interested in, the expansions
\eqref{eq:Th1}--\eqref{eq:Th4} for the Jacobi theta functions
typically require only a small number of terms to converge to within
the machine precision and are therefore well-suited for numerical
evaluation. Using this formulation, we may now compute the Weierstrass
zeta and P functions in \eqref{eq:W-Z2}--\eqref{eq:W-p2} and use them
to evaluate the complex potential and velocity in \eqref{eq:W} and
\eqref{eq:V}.

\section{Computational Approach}
\label{sec:numer}

In this section we present and validate the computational approach
used to evaluate the total drift area \eqref{eq:D2} for different
geometric configurations.  Numerical computation of particle
trajectories and the corresponding drift is performed in a similar
manner to the approach described in detail in \cite{melkoumian2014}.
That is, we solve system \eqref{eq:dxdt} with different initial data
$\x_0 = [0, y_0]^T$, where $|y_0| > 1$. The velocity $\u(\x)$ on the
RHS of \eqref{eq:dxdt} is obtained in the complex form using formula
\eqref{eq:V} {transformed to the fixed frame of reference},
where the Weierstrass zeta and P functions are evaluated using the
Jacobi theta functions as described in Section
\ref{sec:two-cylinderflow}. We then integrate \eqref{eq:dxdt} using
MATLAB's {\tt ode113} routine. Integral \eqref{eq:xi} describing the
total displacement of a particle is an improper one and hence requires
truncation at some suitably large $t = \pm T$. While the numerical
approximation of the integral is declared converged for values of time
$t$ when the velocity magnitude drops below the machine precision
$\epsilon$, i.e., as soon as $|\u(\x(t))| < \epsilon $, to be on the
safe side, we set the maximum integration time $T$ to {\tt realmax}
$=\O(10^{300})$, which is the largest number representable in the
double precision. In order to ensure suitable accuracy of the
time-stepping, the absolute and relative tolerance of the numerical
integration performed by the {\tt ode113} routine were selected as
{\tt RelTol} $=$ {\tt AbsTol} $= 10^{-13}$.

\begin{figure} 
\centering
\mbox{
\begin{subfigure}[b]{0.5\textwidth}
  \includegraphics[width=\textwidth]{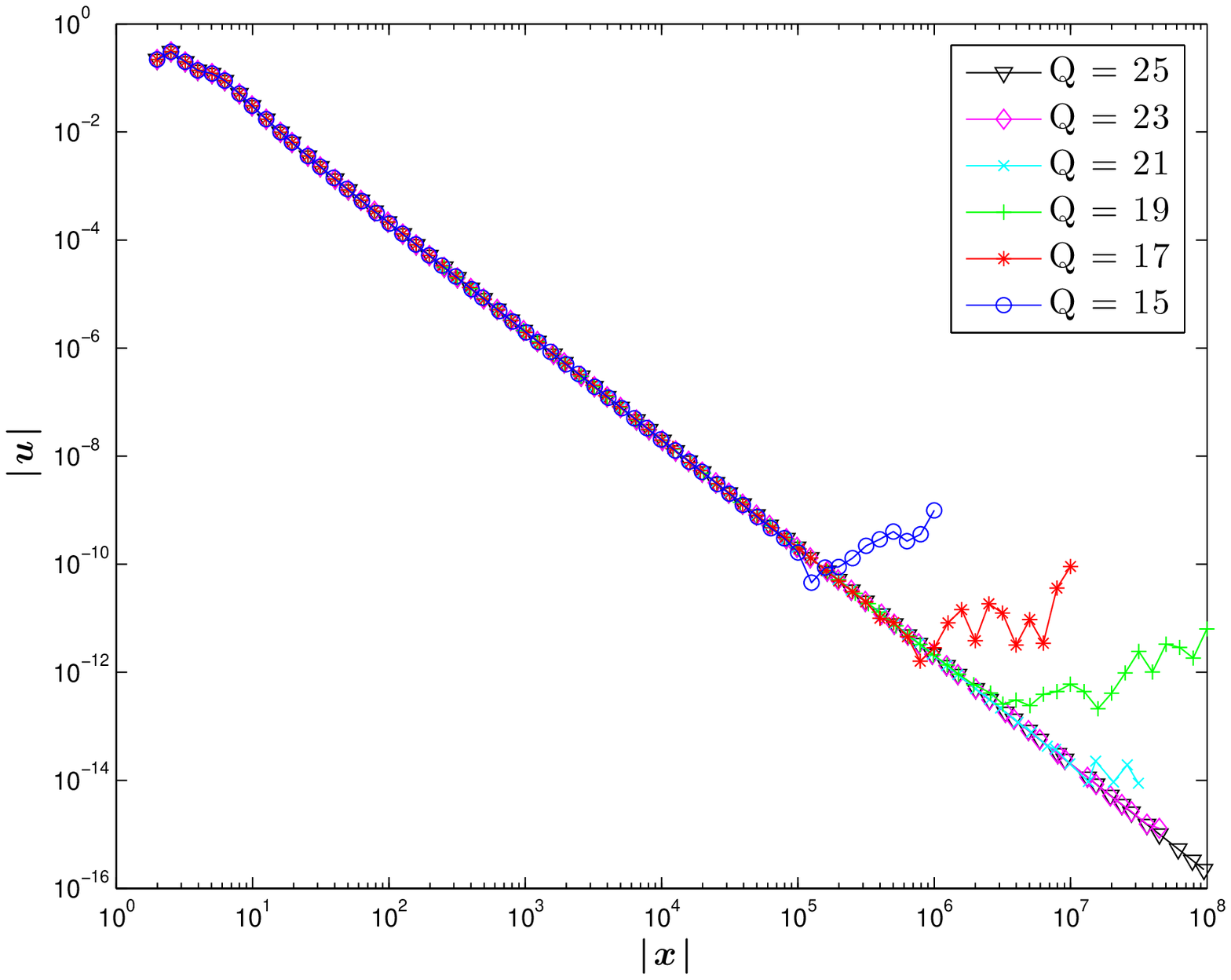}
  \caption{}
	\label{fig:roundoff}
\end{subfigure}
\begin{subfigure}[b]{0.5\textwidth}
  \includegraphics[width=\textwidth]{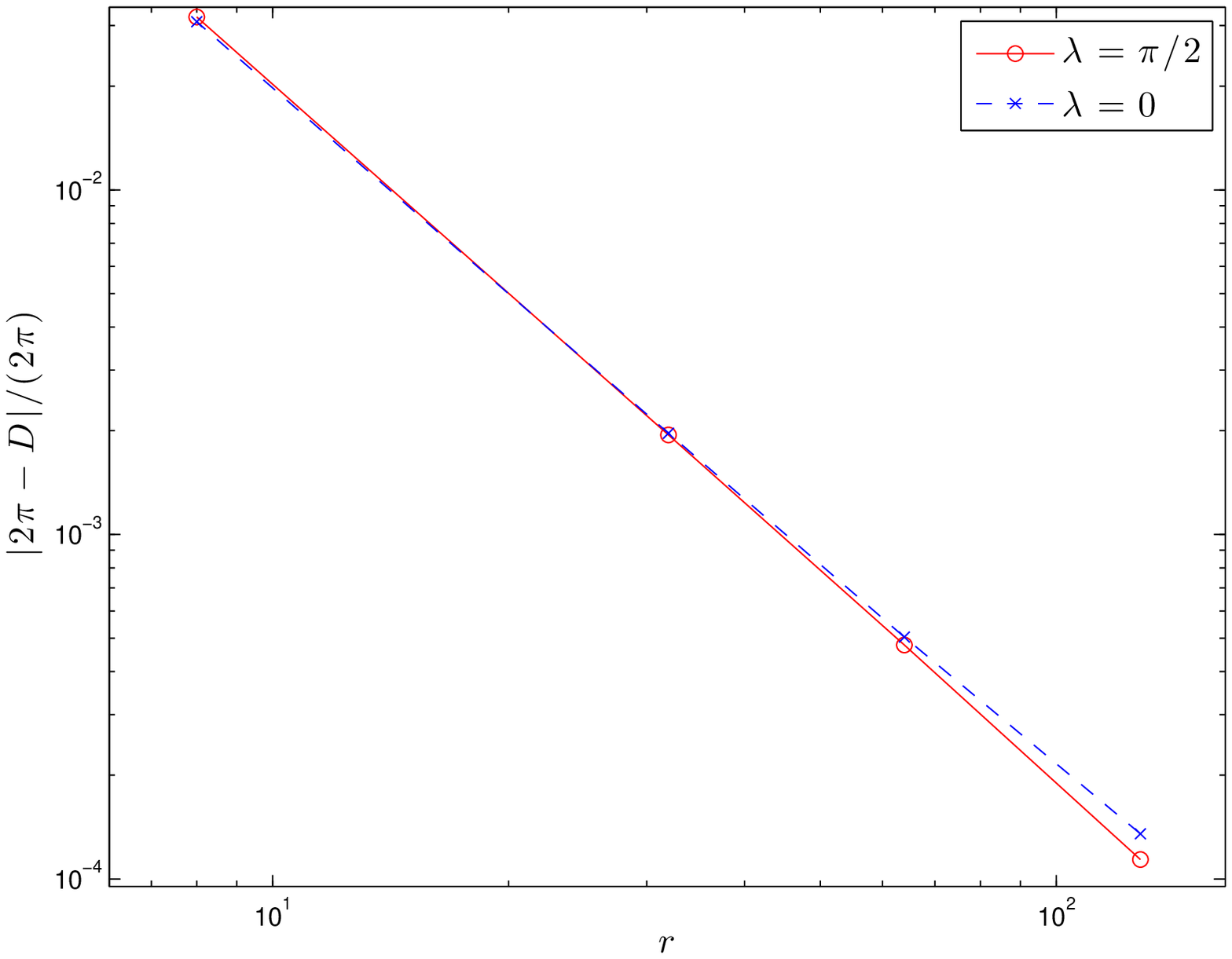}
  \caption{}
	\label{fig:largesep}	
\end{subfigure}
}
\caption{Validation of the computational approach: (a) dependence of
  $|\u([x,0]^T)|$ on $x$ with round-off errors becoming more evident
  as the arithmetic precision is reduced (the numbers of significant
  digits $Q$, used in the computations are indicated in the legend), (b)
  normalized difference between the total drift area induced by two
  cylinders separated by the distance $r$ and twice the drift of a
  single cylinder as a function of the separation $r$.}
\label{fig:valid}
\end{figure}

In computing the velocity $\u(\x)$ there exist round-off errors due to
finite-precision arithmetic. The element of our approach which is the
most susceptible to these errors is the evaluation of the special
functions in \eqref{eq:W-Z2}--\eqref{eq:W-p2}. However, the impact of
these errors can be controlled by performing the computations with an
increased arithmetic precision which in the present study is achieved
by using the Advanpix Multiprecision Computing Toolbox for MATLAB
\cite{mct2015}. The effect of using different arithmetic precisions on
the round-off errors in the evaluation of velocity is illustrated in
Figure \ref{fig:roundoff}, where we show the dependence of
$|\u([x,0]^T)|$ {when the position of the obstacles is fixed at
  $t=0$} on the streamwise coordinate $x$ for $x$ ranging over several
orders of magnitude. From the potential flow theory we know that, in
the absence of circulation, $|\u(\x)| \sim |\x|^{-2}$ as $|\x|
\rightarrow \infty$ which is what is indeed observed in Figure
\ref{fig:roundoff} for intermediate values of $x$.  For large values
of $x$ we detect deviations from this asymptotic behavior due to
round-off errors which are however reduced when the arithmetic
precision is refined. By performing a number of tests it was
determined that the arithmetic precision involving $Q = 25$
significant digits was sufficient to ensure the accuracy of the drift
evaluation required for the present study.

\begin{figure} 
\centering
\mbox{
\begin{subfigure}[b]{0.5\textwidth}
  \includegraphics[width=\textwidth]{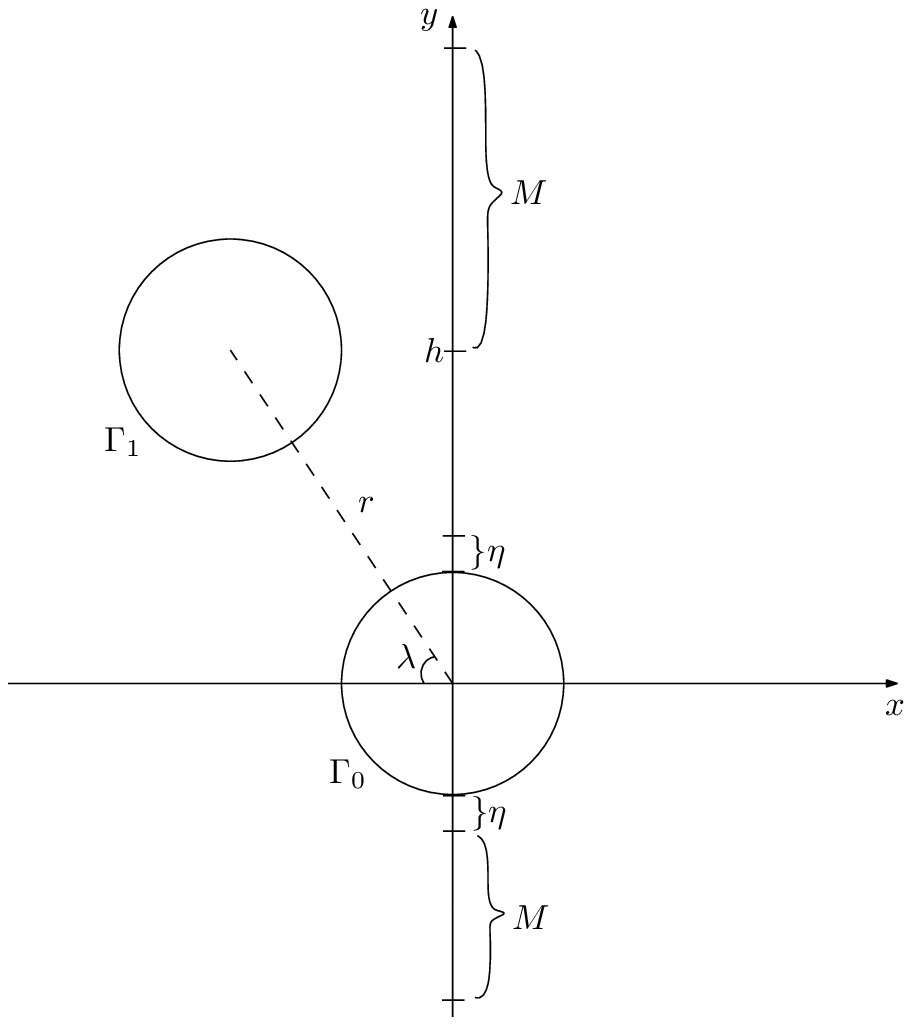}
  \caption{}
	\label{fig:discr1}
\end{subfigure}
\begin{subfigure}[b]{0.5\textwidth}
  \includegraphics[width=\textwidth]{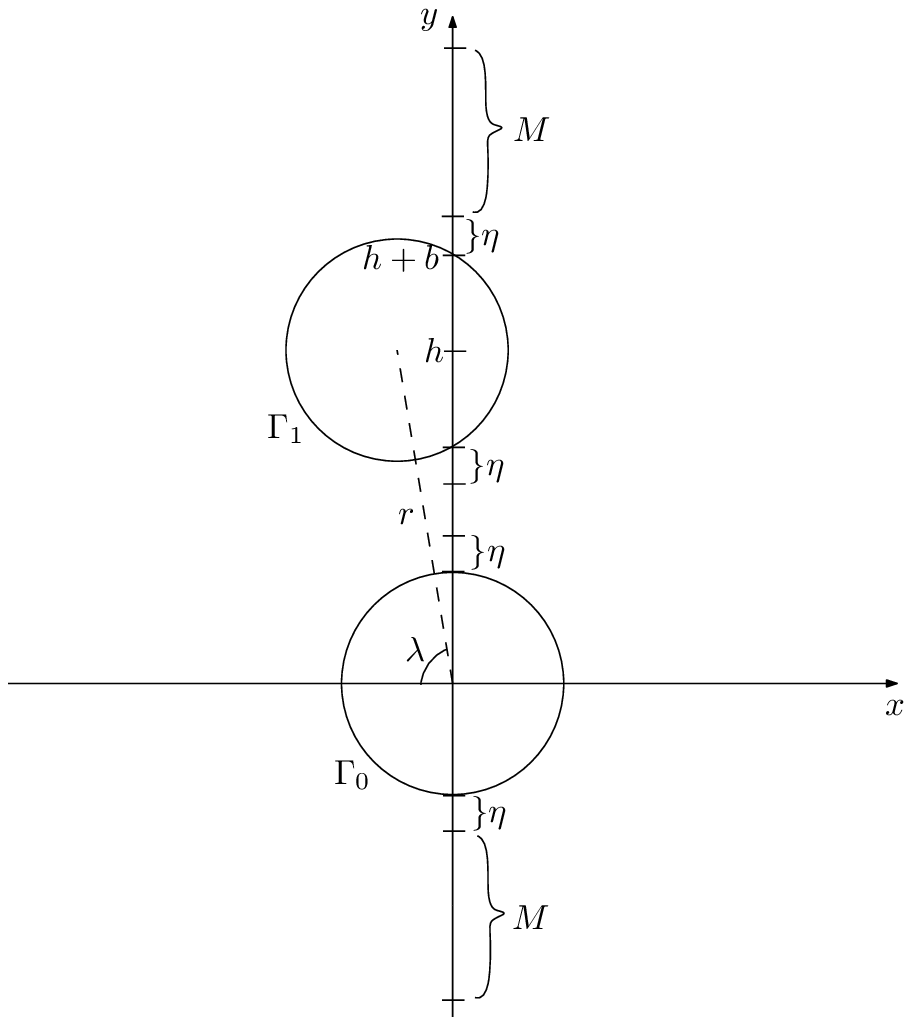}
  \caption{}
	\label{fig:discr2}	
\end{subfigure}
}
\caption{Interpretation of the numerical parameters used in the
  approximation of integrals in \eqref{eq:D2} when the second cylinder
  (a) does not intersect the y-axis and (b) intersects the y-axis
  {at time $t=0$}. The distances corresponding to $\eta$ and $M$
  are not drawn to scale.}
\label{fig:discr}
\end{figure} 

Calculation of the total drift area based on formula \eqref{eq:D2} is
implemented as described below. Since the integrals in \eqref{eq:D2}
are improper, there are two steps required to approximate them, namely,
truncation of the unbounded integration domain (in the variable
$y_0$) and discretization of the resulting definite integrals with
quadratures. As regards the first step, we have the following two
possibilities, depending on the location of the second cylinder (with
the boundary $\Gamma_1$) relative to the y-axis {at time $t=0$}:
\begin{itemize}
\item if the second cylinder {\em does not} intersect the y-axis
  (Figure \ref{fig:discr1}), the integration domain
  $[-\infty,-1]\cup[1,\infty]$ in \eqref{eq:D2} is replaced with
  $[-1-\eta-M,-1-\eta]\cup[1+\eta,h+\eta+M]$, where $h := r
  \sin\lambda$, $M>0$ is a large number and $\eta>0$ is a small number
  (introduced to ensure that the initial particle positions do not
  coincide with the obstacle boundary $\Gamma_0$ which would lead to
  numerical difficulties in solving \eqref{eq:dxdt}),

\item if the second cylinder does intersect the y-axis (Figure
  \ref{fig:discr2}), the integration domain in \eqref{eq:D2} is split
  into {\em three} parts
  $[-1-\eta-M,-1-\eta]\cup[1+\eta,h-b-\eta]\cup[h+b+\eta,h+b+\eta+M]$,
  where $b := \sqrt{1 - r^2 \cos^2\lambda}$.
\end{itemize}
In the computations presented below we used $\eta = 10^{-8}$ and $M =
200$ which were found to ensure the required accuracy of the total
drift area. The definite integrals obtained as a result of this domain
truncation were then approximated using trapezoidal quadratures with
the grid (in $y_0$) selected in such a way that the relative
difference of $\xi(y_0)$, cf.~\eqref{eq:xi}, corresponding to two
adjacent quadrature points did not exceed $1\%$.  To find the function
${\dot{g}}(y_0)$ appearing in \eqref{eq:D2}, we used the
property of potential flows that ${u'_x = \Re{(V)} = \partial
  \psi' / \partial y}$, so that we obtained,
cf.~\eqref{eq:C}--\eqref{eq:dg},
\begin{equation}
{\dot{g}(y_0) = \frac{d y_{\infty}}{d y_0} = \frac{d \psi'(0, y_0)}{d y_0}
= \frac{\partial \psi'(x', y)}{\partial y} \bigg|_{x' = 0, y=y_0}} = \Re{(V(z))} \bigg|_{z = iy_0}
\label{eq:dg2}
\end{equation}
which can be easily evaluated.

To validate the entire numerical approach, we benchmark our
computations against a test case when the cylinder separation $r$ is
very large. In this limit the exact result is known, because when
separated by an infinite distance, the two cylinders do not interact
and the total drift area is twice the drift induced by an individual
obstacle which is $D_0 = \pi$. Thus, we have $D \rightarrow 2 \pi$ as
$r \rightarrow \infty$.  Figure \ref{fig:largesep} shows the results
of this test performed for the tandem ($\lambda = 0$) and transverse
($\lambda = \pi/2$) configurations. It is evident from this plot that the
total drift area indeed approaches $2\pi$ for increasing separations
$r$.

\section{Results} 
\label{sec:results}

In this section we present our computational results, first focusing
on the trajectories of individual particles for different cylinder
configurations and then studying the resulting total drift areas.

\subsection{Individual Particle Trajectories}

\begin{figure}
\vspace*{-1.8cm}
\centering
  \begin{subfigure}{0.6\textwidth}
    \includegraphics[width=\textwidth]{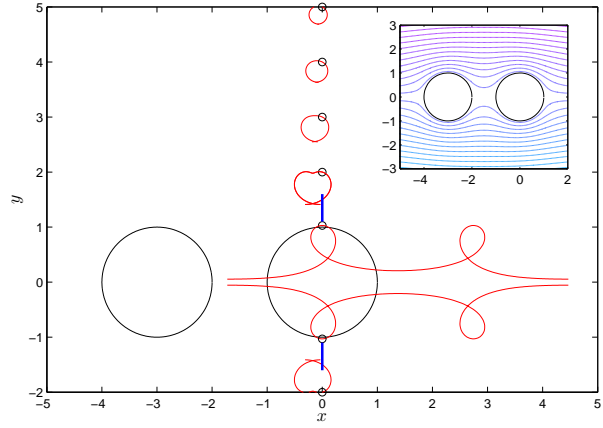} \\ \vspace*{-0.8cm}
    \caption{tandem configuration ($\lambda = 0$)} 
    \label{fig:tandem}
  \end{subfigure}
  \begin{subfigure}{0.6\textwidth}
    \includegraphics[width=\textwidth]{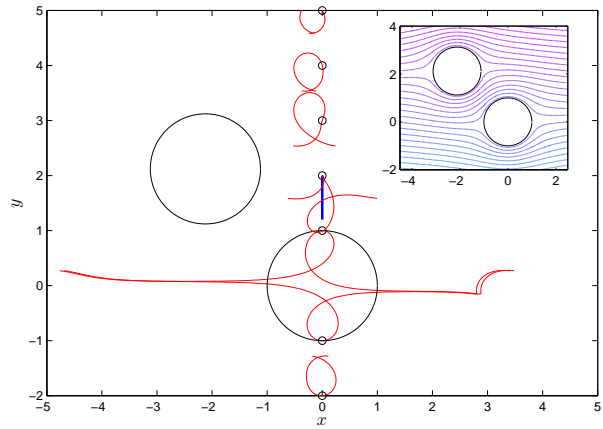} \\ \vspace*{-0.8cm}
    \caption{angled configuration ($\lambda = \pi/4$)}
    \label{fig:angled}
  \end{subfigure}
  \begin{subfigure}{0.6\textwidth}
    \includegraphics[width=\textwidth]{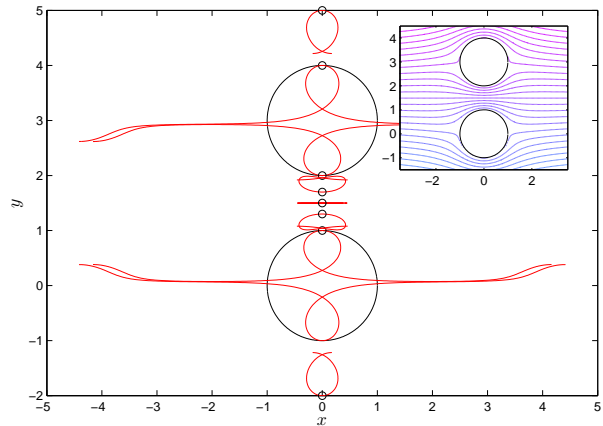} \\ \vspace*{-0.8cm}
    \caption{transverse configuration ($\lambda = \pi/2$)}
    \label{fig:transverse}
  \end{subfigure}
  \caption{Particle trajectories for different initial conditions
    $\x_0=[0,y_0]^T$ in (a) the tandem, (b) the angled and (c) the
    transverse cylinder configurations. The symbols $\circ$ denote the
    particle positions at time $t = 0$ when the cylinders {are
      at the indicated locations.} The blue solid lines in (a) and (b)
    indicate the range of $\x_0$ for which the trajectories exhibit
    two loops, {whereas the insets show the corresponding
      streamline patterns in the moving frame of reference.}}
\label{fig:traj}
\end{figure}

Individual particle trajectories are studied in order to understand
the kinematic mechanisms responsible for the different displacements
the particles undergo. We focus on the trajectories of particles with
initial positions $\x_0=[0,y_0]^T$ in the fixed frame of reference for
three representative cylinder configurations, namely, the tandem
($\lambda = 0$), angled ($\lambda = \pi/4$) and transverse ($\lambda =
\pi/2$) configuration with the cylinder separation $r = 3$ in all
cases. The trajectories corresponding to different values of $y_0$ are
shown in Figures \ref{fig:traj}a,b,c {together with the
  corresponding streamline patterns (in the moving frame of
  reference). Animated versions of these figures are available as
  Online Resource 1, 2 and 3 accompanying this paper.}

In the tandem configuration in Figure \ref{fig:tandem} all particle
trajectories are symmetric with respect to the flow centerline and we
find that the trajectories of the particles passing close to the
cylinders exhibit two well defined loops. These trajectories are
qualitatively similar to the elastica curves describing the particle
trajectories in the single cylinder case \cite{mt1968}. A loop occurs
when the particle changes direction as a result of a cylinder passing
directly above or below it. Since there are two cylinders, for
particles initially close to the flow centerline (i.e., with small
$|y_0|$), we observe two such loops, each associated with the passage
of one cylinder. On the other hand, for particles further away from
the flow axis, the trajectories have only one loop {whose shape}
approaches a circle as $|y_0|$ increases. The range of the initial
positions $\x_0$ for which the trajectories exhibit two loops is
marked with a blue solid line in Figure \ref{fig:tandem} and we find
that this range is confined to a region close to the obstacles.

The trajectories of particles in the angled cylinder configuration are
shown in Figure \ref{fig:angled}. In this case, the vertical symmetry
is broken and the trajectories are clearly more complicated. We find
that there are again some initial positions $\x_0$ such that the
corresponding trajectories exhibit two loops, however, the trajectories
are qualitatively different from those observed in the tandem
configuration (Figure \ref{fig:tandem}). In addition, there is also a
range of initial positions $\x_0$ for which the particles, despite
passing very close to $\Gamma_0$, have trajectories which contain only
one loop.  Furthermore, particles with initial positions close to
$\Gamma_0$ have trajectories exhibiting a ``kink'' at large times
$t>0$. This kink occurs at the {instant} when the second cylinder with
the boundary $\Gamma_1$ passes from above drawing the particle toward
its rear stagnation point. Particles with initial locations further
away from the cylinders again show more circular trajectories.

In the transverse cylinder configuration shown in Figure
\ref{fig:transverse}, the vertical symmetry is now {restored}
with the symmetry axis at $y=1.5$ half-way between the two cylinders.
For a particle initially located at $\x_0 = [0, 1.5]^T$, there is no
vertical displacement, because the vertical components of the velocity
induced by the two cylinders cancel. Regardless of the initial
transverse coordinate $y_0$, all particles follow trajectories with
only one loop which is because their motions are dominated by the
cylinder they are closest to.

\subsection{Total Drift Area}
\label{sec:D}

\begin{figure}
\centering
\includegraphics[width=0.6\textwidth]{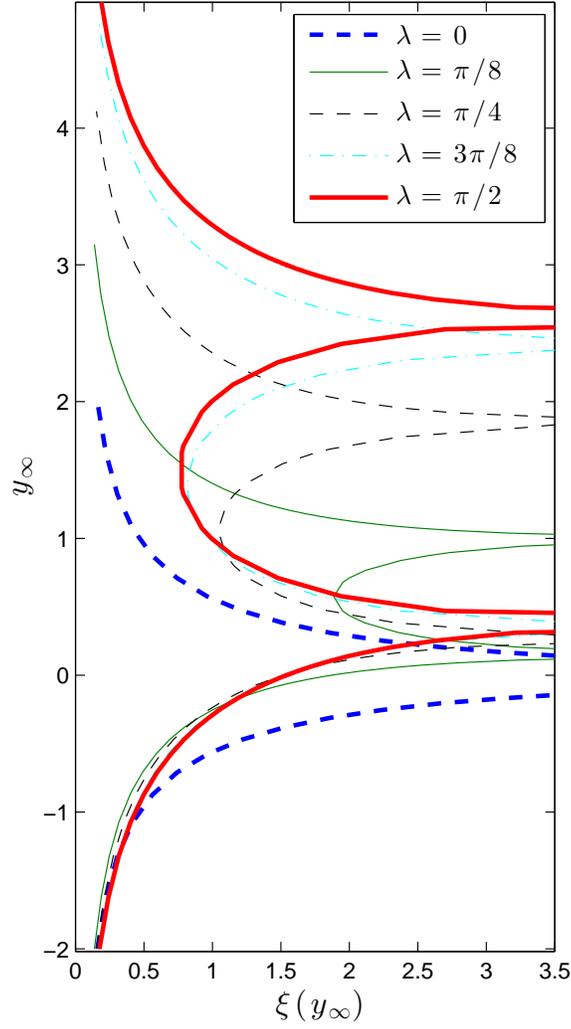}
\caption{Displacement $\xi(y_\infty)$ of particles located initially
  {(i.e., when $t = -\infty$)} at $[0,y_{\infty}]^T$ {as a
    function of $y_{\infty}$} for cylinders with different inclination
  angles indicated in the legend. The separation between the cylinders
  is $r=3$.}
\label{fig:xi}
\end{figure}

In this section we present and analyze some global diagnostic
quantities characterizing the displacement of the particles depending
on the geometric configuration of the two cylinders. We begin by
plotting the displacement $\xi(y_\infty)$ of particles located at
$[0,y_{\infty}]^T$ at time $t = - \infty$ {as a function of
  $y_{\infty}$ for different} inclination angles $\lambda$ in Figure
\ref{fig:xi} (for consistency with the set-up of the problem, the
``independent'' variable $y_{\infty}$ {is measured along} the
vertical axis). The quantity $\xi(y_\infty)$ represents the total
distance travelled by a particle initially at $y_{\infty}$ as the
cylinders move from $x=-\infty$ and $x=\infty$,
cf.~\eqref{eq:C}--\eqref{eq:dg}. {Formation of the profiles
  $\xi(y_\infty)$ during the passage of the obstacles in shown for
  $\lambda = 0, \pi/4, \pi/2$ in Online Resource 1, 2 and 3
  accompanying this paper}.  In Figure \ref{fig:xi} we see that
$\xi(y_\infty)$ becomes unbounded for certain values of $y_{\infty}$
which occurs when the corresponding streamline is connected to the
front stagnation point on one of the cylinders (a phenomenon which is
well understood in the single cylinder case \cite{childress2009}). The
quantity $\xi(y_\infty)$ diverges, respectively, for one and two
values of $y_{\infty}$ when the inclination angle is $\lambda = 0$ or
$0 < \lambda \le \pi/2$, which reflects the number of the front
stagnation points facing the flow. In Figure \ref{fig:xi} we also
observe that as $\lambda$ increases from 0 to $\pi/2$ the distance
between the values of $y_{\infty}$ for which $\xi(y_\infty)$ diverges
increases. We {reiterate} that the total drift area is obtained
by integrating $\xi(y_\infty)$ with respect to $y_\infty$,
cf.~\eqref{eq:D}.

\begin{figure}
\centering
\includegraphics[width=0.6\textwidth]{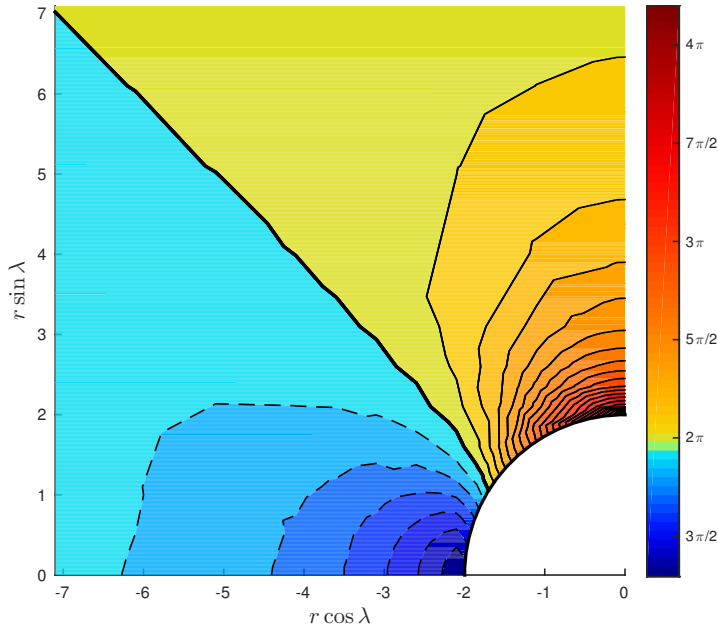}
\caption{Total drift area $D$ as a function of the horizontal and
  vertical separation between the two cylinders. {The thick
    solid line corresponds to $D = 2\pi$, whereas the thin dashed and
    solid lines represent isocontours of $D$, respectively, in areas
    where $D < 2\pi$ and $D > 2 \pi$.}}
\label{fig:Dmap}
\end{figure}

Our main result concerns the dependence of the total drift area $D$ on
the horizontal and vertical separation, respectively $r\cos\lambda$
and $r\sin\lambda$, between the two cylinders and this data is shown
in Figure \ref{fig:Dmap}. Since $r>2$, this region is left blank in
the bottom right corner of the plot. We see that there are two regions
in this parameter space corresponding to $D < 2\pi$ and $D > 2\pi$
where the total drift area is, respectively, decreased or increased
with respect to twice the drift of a single cylinder. This illustrates
how the hydrodynamic interaction of the cylinders due to the geometric
nonlinearity of the problem impacts the displacement of the particles.
In Figure \ref{fig:Dmap} we see that the {total drift area} is
decreased with respect to the non-interacting reference case with $D =
2 \pi$ for small inclination angles $\lambda$ {corresponding to}
more ``streamlined'' {cylinder} configurations, whereas the
opposite effect is observed for larger inclination angles. For
increasing {separation} $r$ the border between the regions with
$D < 2\pi$ and $D > 2\pi$ approaches a straight line described by
$\lambda = \pi / 4$. The deviation of the total drift area from the
reference value $D = 2 \pi$ increases for small separations $r
\rightarrow 2$.

\begin{figure}
\centering
\includegraphics[width=0.7\textwidth]{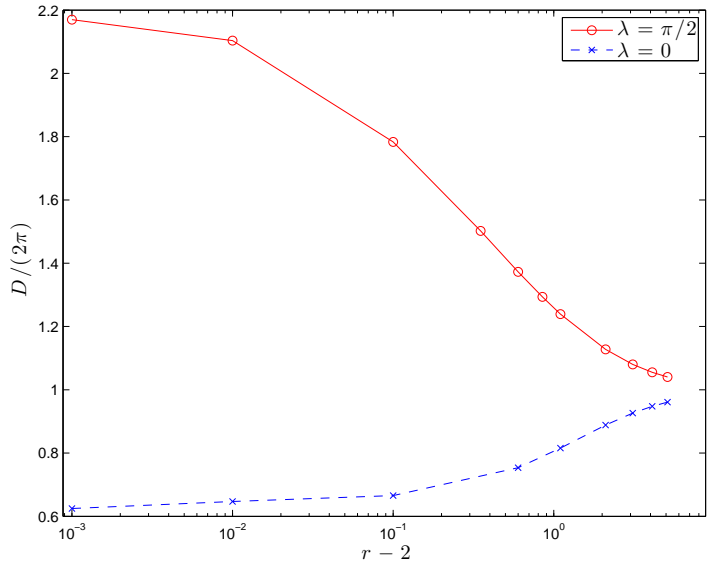}
\caption{Normalized total drift area $D$ as a function of the
  separation $r$ between the cylinders in the tandem {($\lambda
    = 0$)} and transverse {($\lambda = \pi / 2$)}
  configurations.}
\label{fig:minmax}
\end{figure}

It is interesting to find out what are the largest and smallest values
of the total drift area which can be attained. This question is
addressed in Figure \ref{fig:minmax} where we show the values of $D$
for $r \rightarrow 2$ in the tandem ($\lambda = 0$) and transverse
($\lambda = \pi / 2$) configurations ({numerical} evaluation of
the drift when $r = 2$, i.e., when the two cylinder touch, is not
possible due to a singularity of the potential flow formulation,
cf.~Section \ref{sec:two-cylinderflow}). We see that the largest and
smallest drift values are, respectively, $D \approx 2.2 \times 2 \pi$
and $D \approx 0.6 \times 2 \pi$, and are attained when the two
cylinders touch in the transverse and tandem configurations. The
increase of the drift in the former case can be understood as being
due to the two cylinders creating a ``cavity'' in front of them which
traps particles as the cylinders advance. On the other hand, in the
tandem configuration the front stagnation point of the rear cylinder
is shielded by the first cylinder which reduces the trapping effect.

\section{Discussion, Conclusions and Outlook} 
\label{sec:final}

{In this study we have focused on the drift induced by the
  passage of two circular cylinders in an unbounded incompressible
  fluid under the assumptions that the flow is potential and
  stationary in the moving frame of reference. Knowing that when the
  two cylinders are separated by an infinite distance, the total drift
  area is equal to twice the drift induced by one cylinder, the goal
  was to analyze how the geometric configuration of the cylinders,
  namely, their separation and inclination with respect to the
  translation direction, affect the total drift. For each considered
  configuration, the problem was studied by integrating particle
  trajectories numerically with high precision for a range of
  different initial particle positions. The velocity field induced by
  the two cylinders during their passage was expressed in closed form
  in terms of special functions using methods of the complex function
  theory (cf.~Section \ref{sec:models}). In contrast to the
  computation of drift based on Darwin's theorem, the present approach
  provides additional information about the shapes of the particle
  trajectories which sheds light on the kinematic mechanisms
  responsible for the increase or decrease of the total drift area in
  comparison with the reference case when the two cylinders are
  infinitely separated.

  While the set-up of our problem is admittedly highly idealized,
  where we assume an infinite passage time and time-invariance of the
  geometric configuration, this problem nonetheless offers some
  fundamental insights into the relation between the flow geometry and
  drift. We emphasize here that, although the equations governing
  potential flows are linear in the flow variables, they exhibit a
  {\em nonlinear} dependence on the geometry of the flow domain. This
  nonlinearity is manifested in the results reported in Section
  \ref{sec:results}, which demonstrate that the drift induced by
  individual obstacles may not be simply added to produce the total
  drift. More specifically, we showed that for small inclination
  angles $\lambda$, resulting in more ``streamlined'' cylinder
  arrangements, the total drift area is decreased in comparison with
  the reference case, and is increased in configurations characterized
  by the inclination angle $\lambda$ larger than $\pi/4$ (cf.~Figure
  \ref{fig:Dmap}). We also determined the extreme values attained by
  the total drift area $D$ which correspond to the vanishing
  separations between the two cylinders in the tandem and transverse
  configurations (cf.~Figure \ref{fig:minmax}). {We observe
    interesting analogies between these findings and the results
    reported in \cite{lz14} where the authors studied the influence of
    the geometric configuration on the effective diffusivity. It
    appears that both the total drift and the effective diffusivity
    attain their maxima and minima for vanishing separations between
    the obstacles and for inclination angles, respectively, $\lambda =
    \pi / 2$ and $\lambda = 0$. Remarkably, the relative changes of
    the total drift and of the effective diffusivity with respect
    their values in the non-interacting cases were found to be very
    similar --- they are given by the factors of approximately 2.2 and
    0.6 for $\lambda = \pi / 2$ and $\lambda = 0$. On the other hand,
    unlike the total drift area, the effective diffusivity seems to
    converge to twice its value in the non-interacting case as the
    separation distance $r$ increases only for certain inclination
    angles $\lambda$ \cite{lz14}.}  As regards the particle
  trajectories obtained in the different configurations, we observed
  that high-drift cases lead to trajectories with a single loop
  (cf.~Figure \ref{fig:traj}a), whereas in the low-drift
  configurations particle trajectories starting at certain initial
  locations may actually exhibit two loops (cf.~Figures
  \ref{fig:traj}b,c).  {These observed trajectories
    qualitatively match those reported in \cite{lz14}.}  The results
  obtained in the present study may help quantify and improve the
  accuracy of models used to describe {transport and mixing}
  caused by multiple objects in various biological and multiphase flow
  applications.  }

\section*{Acknowledgements}

{The authors wish to thank Prof.~Takashi Sakajo and Dr.~Rhodri
  Nelson for their advice concerning potential flow models for flows
  past multiple obstacles. Partial support for this research was
  provided by an NSERC (Canada) Discovery Grant.}


\end{document}